\pdfoutput=1

\documentclass[11pt]{article}

\usepackage{EMNLP2022}

\usepackage{times}
\usepackage{latexsym}

\usepackage[T1]{fontenc}

\usepackage[utf8]{inputenc}

\usepackage{microtype}

\usepackage{inconsolata}

\usepackage{multirow}
\usepackage{csquotes}
\usepackage{booktabs}
\usepackage{amsmath}
\usepackage{graphicx}

\title{Retrieval Augmentation for T5 Re-ranker using External Sources}

\author{
\textbf{Kai Hui}\thanks{~~Corresponding Author} \quad
\textbf{Tao Chen} \quad 
\textbf{Zhen Qin}  \quad 
\textbf{Honglei Zhuang} \quad\\
\textbf{Fernando Diaz} \quad
\textbf{Michael Bendersky} \quad
\textbf{Donald Metzler} \quad  \\\\ kaihuibj@google.com \\ Google Research }

\begin{document}
\maketitle
\begin{abstract}
Retrieval augmentation has shown promising improvements in different
tasks. 
However, whether such augmentation can assist a large language model based re-ranker remains unclear.
We investigate how to augment T5-based re-rankers using high-quality information retrieved from two external corpora --- a commercial web search engine and Wikipedia.  
We empirically demonstrate how retrieval augmentation can substantially improve the effectiveness of T5-based re-rankers for both in-domain and zero-shot out-of-domain re-ranking tasks.
\end{abstract}
\newcommand{\query}[0]{q}
\newcommand{\doc}[0]{d}
\newcommand{\expansionTerms}[0]{\tilde{\query}}
\newcommand{\sequence}[0]{\mathbf{x}}
\newcommand{\rel}[0]{y}
\newcommand{\score}[0]{\tilde{\rel}}
\newcommand{\snippet}[0]{\sigma}
\newcommand{\eRetrieval}[0]{f}

\section{Introduction}\label{sec.intro}

Large language models (LLMs) 
have demonstrated strong performance for both retrieval and re-ranking tasks
~\cite{qiao2019understanding,nogueira2020T5ranking,lin2021pretrained,tay2022transformer}.
Meanwhile, recent research has enhanced LLMs by augmenting them with retrieval capabilities. These retrieval augmentation strategies have demonstrated promising results on tasks such as
question answering~\cite{guu2020realm,lewis2020retrieval,borgeaud2021improving}
and knowledge-grounded conversation~\cite{shuster2021retrieval,cohen2022lamda}.
However, it remains an open question as to whether retrieval augmentation can improve the effectiveness of LLM-based re-rankers.

Inspired by the uses of the external expansions~\cite{diaz2006improving},
we explore how query representations can be augmented using information retrieved from high-quality corpora to improve re-ranking quality.
Similar to RAG~\cite{lewis2020retrieval} and RETRO~\cite{borgeaud2021improving},
we augment a given query with information retrieved from 
external resources before performing inference (i.e., re-ranking in this case).
We consider external resources from 
two corpora --- a commercial web search engine and Wikipedia. 
We consider two approaches to augment the original query with retrieved information. The first approach augments the query with a \emph{sequence} of terms while the second augments the query with a \emph{bag of words}.
Finally, we investigate the efficacy of query augmentation strategies 
for in-domain and zero-shot out-of-domain settings.
We evaluate these conditions on three datasets: open-domain Natural Questions~\cite{kwiatkowski2019natural},
MS MARCO passage re-ranking~\cite{nguyen2016ms}, and TREC DL Track 2019~\cite{craswell2020overview} and 2020~\cite{craswell2021overview}.

This paper makes three contributions: (1) To the best of our knowledge, this is the first investigation of retrieval augmentation for LLM-based re-rankers;
(2) Retrieval augmentation is empirically analyzed using four T5-based reranker variants for in-domain and zero-shot out-of-domain tasks;
(3) A comprehensive experimental study is presented using two retrieval corpora and two different query augmentation strategies. 

\section{Related Work}~\label{sec.related_work}

\noindent\textbf{Retrieval augmentation for NLP tasks.}
Due to the opaque nature of knowledge stored in the parameters of LLMs,
retrieval augmentation has been introduced for a variety of different NLP tasks.
For example, on question answering tasks, REALM~\cite{guu2020realm}, RAG~\cite{lewis2020retrieval} augment inputs with a document corpus, enriching the representation using top-ranked retrieved items via Maximum Inner Product Search (MIPS). 
Meanwhile, RETRO~\cite{borgeaud2021improving} uses retrieval to augment at the granularity of small chunks of tokens.
It has also been shown that retrieval augmentation can help 
provide better-grounded text
in dialogue systems~\cite{shuster2021retrieval,cohen2022lamda} 
and in the evaluation of hallucination~\cite{honovich2021q2}.
Inspired by these successes, our work 
investigates retrieval augmentation for re-ranking using a fixed retrieval component.

\noindent\textbf{Query expansion and pseudo-relevance feedback (PRF).}
In early work,
\citet{diaz2006improving}~showed it is effective to incorporate information from an external corpus into a non-neural language modeling framework.
We exploit such information when using a pre-trained language model for re-ranking
by directly augmenting the original query with the top-ranked results from an external corpus.   
An orthogonal research direction is to improve re-ranking models by
incorporating pseudo-relevance feedback (PRF) signals
as in~\cite{li2018nprf, padaki2020rethinking, zheng2020bert, yu2021pgt,naseri2021ceqe}. One essential component therein identifies the relevant information from the pseudo relevance, avoiding the topic shift. Besides, these methods are involved with expensive multiple iterations to collect the PRF and use that for re-ranking. 
In contrast, our model consumes high-quality external augmentation text and 
requires one single iteration.

\section{Method}\label{sec.method}
We adopt \citeauthor{nogueira2019doc2query}'s method for re-ranking with LLMs \cite{nogueira2019doc2query}.  Let $\query$ be the query string, $\doc$ be the document string, and $\rel$ be a string that represents the binary relevance of a document, e.g., ``True'' or ``False''.  We construct a (string) instance $\sequence$ as,
\begin{align}
    \sequence &= \texttt{"Query:}\query\texttt{ Document:}\doc\texttt{ Relevant:}\rel\texttt{"}
\end{align}
The model is trained to generate the final
token (i.e. $\rel$) based on the ground-truth relevance of the query-document pair.  To score a new query-document pair, the normalised score of the final token is used for re-ranking.

We are interested in augmenting $\sequence$ with information from an external corpus.  We assume that access to the external corpus is mediated through a retrieval service $\eRetrieval$ such that $\eRetrieval(\query)=[\snippet_1,\ldots,\snippet_m]$, where $\snippet_i$ is a retrieved passage (e.g. web search snippet, indexed passage).
It is important to note that the retrieval service can only retrieve items from a given external corpus and cannot re-rank or re-score documents in the target corpus.

We represent the information $\eRetrieval(\query)$ as an augmenting string $\expansionTerms$.  We can directly concatenate the $m$ passages to construct $\expansionTerms$; we refer to this as \textit{natural language expansion}.  Although we expect the natural language expansion to be more compatible with LLMs, the fixed capacity of LLM modeling can result in situations where informative text is obscured by `linguistic glue' often discarded as stop words~\cite{tay2020efficient}.
Alternatively, we can extract the most salient topical terms from $\eRetrieval(\query)$ as in ~\cite{dang2013term}.  Specifically, we select terms using the KL2 method~\cite{carpineto2001information, amati2003probabilistics}. In this method,
we select $k$ terms from all of the terms in $\eRetrieval(\query)$ using each individual words' contribution in the KL-divergence between the language model in $\eRetrieval(\query)$ (denoted as $\mathcal{A}$) and the corpus (denoted as $\mathcal{C}$).
\begin{align}
    w(t, \mathcal{A}) &= P(t|\mathcal{A}) \mathit{log}_2 \frac{P(t|\mathcal{A})}{P(t|\mathcal{C})}\label{eq.kl2}
\end{align}
We estimate the corpus language model using the target retrieval dataset. We refer to this as \textit{topical term expansion}.
In both expansion methods, we truncate the concatenated snippets, paragraphs, or ordered set of topical words (according to Eq.~\ref{eq.kl2}) to a maximum sequence length.

To incorporate retrieved information, represented as $\expansionTerms$ (the expansion terms), we add the text as a new subsequence (``Description'') in $\sequence$,
\begin{align*}
    \sequence = &\texttt{"Query:}\query\texttt{ Description:}\expansionTerms\\
    &\texttt{ Document:}\doc\texttt{ Relevant:}\rel\texttt{"}
\end{align*}
Because we are representing instances as strings with a terminal relevance label, we can easily adopt the same re-ranking method as \citet{nogueira2019doc2query}.  

\section{Experiments}\label{sec.setting}
\begin{table*}[!t]
    \centering
   \caption{The models are fine-tuned on NQ training data and tested
   on NQ-test (in-domain setting) and TREC DL 2019 and 2020 (out-of-domain setting).
    }
    \resizebox{0.95\textwidth}{!}{
    \begin{tabular}{r|c||cccc||cc|cc}
    \toprule
         \multirow{2}{*}{Models}& \multicolumn{1}{c||}{NQ dev} & \multicolumn{4}{c||}{NQ test} &\multicolumn{2}{|c}{TREC DL Track 2019}&\multicolumn{2}{|c}{TREC DL Track 2020}\\
        &S@20& S@1 & S@5 & S@10 &S@20  & nDCG@10& MAP&  nDCG@10& MAP\\
    \midrule
Init ranking &-&52.47	&72.24&77.7&81.33&-&-&-&-\\
    \midrule
T5-small&84.97&50.55&72.33&78.84&83.30&50.36&32.09&52.16&31.27\\
+serp-snippet&88.80 $\Uparrow$&59.64 $\Uparrow$&76.29 $\Uparrow$&81.08 $\Uparrow$&84.35 $\Uparrow$&60.05 $\Uparrow$&37.79 $\Uparrow$&59.19 $\Uparrow$&36.70 $\uparrow$\\
+serp-terms&88.10 $\Uparrow$&58.45 $\Uparrow$&75.15 $\Uparrow$&79.81 $\Uparrow$&83.71 &58.20 $\Uparrow$&36.53 $\uparrow$&54.96 &34.57  \\
\midrule
T5-base&88.69&58.89&77.20&81.08&84.10&57.21&38.98&61.30&37.60\\
+serp-snippet&89.04 &61.41 $\Uparrow$&77.84 &81.69 &84.60 &60.23 &38.96 &57.49 &34.91\\
+serp-terms&89.76 $\Uparrow$&63.74 $\Uparrow$&78.25 $\Uparrow$&82.13 $\Uparrow$&84.43 &63.10 $\uparrow$&40.97 &60.98 &38.70 \\
\midrule
T5-large&90.12&64.46&79.53&82.77&85.15&65.55&42.39&67.87&45.08\\
+serp-snippet&90.50 $\Uparrow$&67.67 $\Uparrow$&80.06 &83.46 $\Uparrow$&85.43 &68.61 &44.27 &64.89 &42.76 \\
+serp-terms&90.65 $\Uparrow$&67.29 $\Uparrow$&79.92 &83.38 $\uparrow$&85.46 &67.63 &43.35 &67.80 &44.43 \\
\midrule
T5-xl&90.56&65.65&80.19&83.46&85.51&67.79&44.06&65.95&42.87\\
+serp-snippet&90.62 &68.28 $\Uparrow$&81.25 $\Uparrow$&83.93 &85.65 &72.12 $\Uparrow$&45.31 &70.32 &47.32 \\
+serp-terms&90.71 &69.64 $\Uparrow$&81.02 $\Uparrow$&83.85 &85.73 &67.46 &43.56 &67.77 &43.80 \\
\bottomrule

    \end{tabular}}
    \label{tab.nq}
\end{table*}
\begin{table*}[!t]
    \centering
  \caption{The models are fine-tuned on MS MARCO training data and tested
  on TREC DL Track 2019 and 2020 (in-domain setting) and NQ-test (out-of-domain setting).
    }
    \resizebox{0.95\textwidth}{!}{
    \begin{tabular}{r|c||cc|cc||cccc}
    \toprule
         \multirow{2}{*}{Models} & \multicolumn{1}{c||}{MSM dev} &\multicolumn{2}{c}{TREC DL Track 2019}&\multicolumn{2}{|c||}{TREC DL Track 2020}&\multicolumn{4}{c}{NQ-test}\\
        & MRR@10 & nDCG@10 & MAP& nDCG@10 & MAP & S@1 & S@5 & S@10 &S@20 \\
    \midrule
Init ranking &18.13&-&-&-&-&52.47	&72.24&77.7&81.33 \\
\midrule
T5-small&35.14&70.83&44.22&68.70&45.13&42.19&69.39&77.09&82.60\\
+serp-snippet&37.12 $\Uparrow$&71.52 &46.63 $\Uparrow$&71.42 $\uparrow$&47.26 $\uparrow$&50.91 $\Uparrow$&74.32 $\Uparrow$&79.75 $\Uparrow$&83.88 $\Uparrow$\\
+serp-terms&36.79 $\Uparrow$&72.09 &45.71 $\Uparrow$&70.81 $\uparrow$&47.29 $\Uparrow$&50.47 $\Uparrow$&73.91 $\Uparrow$&79.89 $\Uparrow$&83.63 $\Uparrow$\\
+wiki-para&36.14 $\Uparrow$&70.04 &45.21 $\uparrow$&68.38 &45.89 &-&-&-&-\\
+wiki-terms&35.55 &70.22 &44.67 &69.58 &46.08 &-&-&-&-\\
\midrule
T5-base&38.50&71.27&45.83&72.50&48.59&46.29&72.77&79.39&83.55\\
+serp-snippet&39.29 $\Uparrow$&72.12 &46.76 &74.68 $\uparrow$&50.06 &51.58 $\Uparrow$&75.54 $\Uparrow$&80.50 $\Uparrow$&84.10 $\Uparrow$\\
+serp-terms&39.25 $\Uparrow$&73.36 &47.25 &73.29 &50.06 $\uparrow$&51.69 $\Uparrow$&75.37 $\Uparrow$&81.27 $\Uparrow$&84.27 $\Uparrow$\\
+wiki-para&38.90 &72.93 &46.05 &73.38 &49.24 &-&-&-&-\\
+wiki-terms&38.51 &72.14 &46.75 &72.72 &49.14 &-&-&-&-\\
\midrule
T5-large&39.91&71.65&46.22&72.65&49.82&49.97&74.49&80.72&83.96\\
+serp-snippet&40.59 $\Uparrow$&72.14 &47.39 $\Uparrow$&75.49 $\Uparrow$&51.57 $\Uparrow$&54.76 $\Uparrow$&77.42 $\Uparrow$&81.88 $\Uparrow$&84.54 $\Uparrow$\\
+serp-terms&40.45 $\Uparrow$&72.27 &47.51 $\Uparrow$&75.54 $\Uparrow$&51.87 $\Uparrow$&53.96 $\Uparrow$&76.84 $\Uparrow$&81.88 $\Uparrow$&84.54 $\Uparrow$\\
+wiki-para&40.34 $\uparrow$&71.42 &46.74 &73.85 &50.65 &-&-&-&-\\
+wiki-terms&40.34 $\uparrow$&72.14 &47.45 $\Uparrow$&73.28 &49.98 &-&-&-&-\\
\midrule
T5-xl&40.15&71.57&46.06&73.59&50.70&51.52&76.20&81.30&84.60\\
+serp-snippet&41.06 $\Uparrow$&73.20 $\uparrow$&47.25 $\uparrow$&76.16 $\Uparrow$&52.41 $\Uparrow$&55.29 $\Uparrow$&77.67 $\Uparrow$&82.35 $\Uparrow$&84.85 \\
+serp-terms&40.83 $\Uparrow$&72.99 $\uparrow$&47.51 $\Uparrow$&76.39 $\Uparrow$&53.49 $\Uparrow$&55.73 $\Uparrow$&78.17 $\Uparrow$&82.24 $\Uparrow$&84.85 \\
+wiki-para&40.59 $\uparrow$&72.78 &47.22 $\Uparrow$&75.24 $\uparrow$&51.80 &-&-&-&-\\
+wiki-terms&40.67 $\Uparrow$&72.70 &47.66 $\Uparrow$&76.49 $\Uparrow$&52.52 $\Uparrow$&-&-&-&-\\
\bottomrule
    \end{tabular}}
    \label{tab.msmarco}
\end{table*}
\paragraph{Training data.}
We use two training datasets, namely,
 Natural Questions (NQ) originally proposed in~\cite{kwiatkowski2019natural}, and,
the MS MARCO~\cite{nguyen2016ms} 
passage re-ranking dataset.
The NQ dataset includes 79k user queries
 from the Google search engine. 
The subset of NQ derived in~\cite{karpukhin2020dense}
are used. The data has the form (question, passage, label), where only the queries with short answers are included.
The task is to retrieve and re-rank the chunked paragraphs
 from Wikipedia with up to 100 words for the queries.
Meanwhile, we use the MS MARCO triplet training dataset~\cite{nguyen2016ms}, which 
 includes 550k positive query-passage pairs.
 For validation purposes, we measure Success@20 (also called Hits@20) on the 8757 questions in the NQ dev dataset, and  MRR@10 on the 6980  queries from the MS MARCO small development dataset.
The training data from both datasets are balanced by up-sampling the positive training samples.

\paragraph{Test data.}
We report Success@1, 5, 10, and, 20 (short for S@x) by re-ranking the top-100 search results for the 3610 test questions
from DPR~\cite{karpukhin2020dense} (the run named adv-hn)\footnote{\url{https://github.com/facebookresearch/DPR}}
as the in-domain test results for models trained on NQ and zero-shot results for models trained on MS MARCO.
We report nDCG@10 and MAP on TREC DL 2019~\cite{craswell2020overview} and 2020~\cite{craswell2021overview} as the in-domain test results for models trained on MS MARCO and zero-shot results for models trained on NQ.
The initial BM25 rankings for MS MARCO are generated 
using Terrier~\cite{DBLP:conf/sigir/MacdonaldMSO12}.

\paragraph{Augmentation details and model variants.}
We fine-tune T5-small, base, large, and, xl models on NQ and MS MARCO,
with and without augmentation.
Specifically, 
we investigate the individual expansions
described in Section~\ref{sec.method}.
For the search engine snippets, we collect up to five organic search result snippets per query.
using a commercial search engine.
The direct answers that appear in the top of the search results, if any, for answer-seeking queries are skipped to avoid the leaking of the answers in the augmentation.
For the  Wikipedia paragraphs, we use non-overlapping text chunks with up to 100 words from DPR~\cite{karpukhin2020dense} for expansion.
The hybrid method~\cite{lin2021few,ma2021replication} from Pyserini\footnote{\url{https://github.com/castorini/pyserini}} ($alpha=1.3$) is used to gather the initial top-1k Wiki paragraphs. 
Thereafter, a T5-xl re-ranker is fine-tuned on the NQ dataset to identify the most relevant paragraphs.
Under the natural language setting, we truncate the augmentation text to 64 words, which is about two web search snippets or one Wikipedia passage.
Under the topical term setting, we use the top-64 topical terms from the top-5 web search snippets or Wikipedia passages.
Statistical significance relative to the non-augmented T5-based rerankers is reported using a two-tailed paired $t$-test.
For the models trained on NQ, we only report the augmentation results using web search snippets. 
This is because we exploit the Wikipedia corpus pre-processed from the DPR paper~\cite{karpukhin2020dense},
leading to the same corpus as in our NQ benchmark,
making the augmentation non-external and is not something that we are exploring here. 

\paragraph{Training.}
  We employ Mesh Tensorflow~\cite{shazeer2018mesh}
for training and evaluation.
All of the T5 models are initialised with pre-trained 
checkpoints. 
The models are fine-tuned and inferred as 
in~\cite{nogueira2020T5ranking}. Instead of selecting models from a fixed step, all models are trained up to 150k steps and
are selected according to their validation set performance. 
The query expansion model is trained
by augmenting inputs as described in Section~\ref{sec.method}. 
During training, a constant learning rate of $1e\text{-}3$ is used for all models.

\section{Results}\label{sec.results}
We summarize results for the NQ and MS MARCO models in Table~\ref{tab.nq} and~\ref{tab.msmarco}, respectively, with validation performance in the second column.
We also include the performance of the initial ranking for reference.

For the \textit{in-domain experiments}, we observe improvements both over the initial ranking and the non-augmented model
on both NQ and MS MARCO when augmenting queries with external information. 
In the NQ development and test sets (Table \ref{tab.nq}), 
expanding using search results yields consistent boosts (more than 4\% relative improvement in S@1 for T5-base, T5-large, and T5-xl models; more than 15\% improvements for T5-small).
We observe similar behavior for the MS MARCO development set (Table \ref{tab.msmarco}), where the use of search engine result augmentation yields consistent boosts.
Furthermore, despite the small number of test queries in TREC DL Track 2019 and 2020, augmentation provides consistently significant improvements across
different T5 variants.

For the \textit{out-of-domain experiments},
 we measure the zero-shot effectiveness of models trained on a different dataset.
 When training on the NQ dataset (Table~\ref{tab.nq}), 
the models, with or without augmentation, fall short on TREC DL tracks. 
This could be due to the difference in tasks (QA in NQ versus IR in TREC DL) or the relatively small amount of training data available in NQ ($\sim$50k unique queries).  
However,
significant improvements are still observed for T5-small on TREC DL Track 2019 and 2020.
By comparison, when training with MS MARCO ($\sim$550k unique queries), the zero-shot performance on NQ is consistently improved over S@1, S@5, and S@10 across different model variants.
Noticeably, no non-augmented T5 variant outperformed the initial DPR ranking in terms of S@1,
while augmented T5-large and T5-xl outperformed the initial ranking.
These findings highlight the potential value  retrieval augmentation has to improve re-ranking quality in out-of-domain settings. 

\textit{Augmentation for different T5 model variants.}
From Table~\ref{tab.nq} and~\ref{tab.msmarco}, it can be seen that augmentation yield larger gains for smaller models. 
For T5-small and base, augmentation tends to provide
more consistent improvements for both in-domain and out-of-domain settings;
whereas for T5-large and T5-xl, 
augmentation helps in the in-domain setting, but 
does not always help in the out-of-domain setting. For example, augmentation barely helps on TREC DL when being trained on NQ, and diminishing improvements are observed in terms of S@20 on NQ-test for models trained on MS MARCO.

With \textit{different sources and strategies for augmentation}, 
using the MS MARCO development set (Table~\ref{tab.msmarco}) as an example, augmentation from both sources improved performance across different T5 variants. 
Information from Wikipedia resulted in smaller improvements compared to search engine information, suggesting that either corpus size (or comprehensiveness) or representativeness (e.g., domain difference) may be important when selecting an augmentation source.
Furthermore, across experimental conditions, improvements from
natural language and topical words were comparable, without strong evidence of one being superior.

Relationship with \textit{pseudo-relevance feedback (PRF)}.
As mentioned in Section~\ref{sec.related_work}, the use of PRF is orthogonal, thus we do not include PRF models for comparisons. Therein, a carefully-designed component to avoid topic shift is crucial for PRF model to perform,
which is not something that we are investigating here.
Actually,
\citet{padaki2020rethinking} demonstrated that PRF on top of BERT-based re-ranker reduced ranking quality by more than 10\%, e.g., nDCG@10 decreases from 0.498 to 0.448 after using PRF expansion (namely, ``ClassicQEConcepts'') as reported in Table 3; and the direct use of suggested questions from Google as query expansion can barely help, getting 0.502 of nDCG@10 (namely, ``GoogleQuestions''); unless these questions are manually filtered, achieving nDCG@10=0.526 using ``FilteredGoogleQuestions''.
Herein, the use of ``FilteredGoogleQuestions'' gets close to our settings, but with different resources and involving manual filtering in extra.

\section{Conclusion}\label{sec.conclusion}
We investigated the use of augmentation from high-quality external sources
for re-ranking tasks, using two external sources, namely, snippets from
a commercial search engine and Wikipedia passages. We also investigated representing the external information as natural language sequences and as bags of words.
We demonstrated the effectiveness of retrieval augmentation for T5 re-rankers
for both in-domain and zero-shot out-of-domain settings.

\section{Limitations}~\label{sec.limitation}
As a showcase of retrieval augmentation for T5-based re-rankers, 
this work did not comprehensively investigate models beyond T5, like duoT5~\cite{pradeep2021expando},
and did not explore more advanced modeling of the external sources, like co-training the augmentation components 
together with the re-ranker as in REALM~\cite{guu2020realm}, RAG~\cite{lewis2020retrieval} and RETRO~\cite{borgeaud2021improving}.
Applying established methods on top of the retrieval augmented re-ranker to further advance
the state-of-the-art could be another research direction. For example, we could have applied pseudo relevance feedback (PRF) on top of the retrieval augmentation re-ranker to see if there exists further improvements. 
Finally, an orthogonal but highly-related direction is to use  retrieval augmentation to improve retrieval, which is not something that we explored here. Instead, this work assumes a fixed retriever that does not use augmentation is used.

\bibliography{ranking}
\bibliographystyle{acl_natbib}
\newpage
\appendix

\end{document}